\documentclass[aps,amssymb,amsmath,prl,twocolumn, showpacs, superscriptaddress]{revtex4}
\usepackage{graphicx}
\usepackage{dcolumn}
\usepackage{bm}

\usepackage{amssymb,amsmath,amsfonts,latexsym,graphicx,verbatim}
\usepackage[matrix,frame,arrow]{xy}
\usepackage{epsf}

\begin{document}
\title{Svetlichny's inequality and genuine tripartite nonlocality in
three-qubit pure states}
\author{Ashok Ajoy}
\email{ashok.ajoy@gmail.com}
\affiliation{Birla Institute of Technology and Science - Pilani, Zuarinagar, Goa - 403726, India.}
\affiliation{NMR Research Centre,
Indian Institute of Science, Bangalore - 560012, India.}
\author{Pranaw Rungta}
\affiliation{NMR Research Centre,
Indian Institute of Science, Bangalore - 560012, India.}
\affiliation{IISER Mohali, Sector-26 Chandigarh - 160019, India.}



\begin{abstract}
The violation of  the Svetlichny's inequality (SI) [Phys. Rev.  D {\bf 35}, 3066 (1987)] is sufficient but not necessary for genuine tripartite nonlocal correlations. Here we quantify the relationship between tripartite entanglement and the maximum expectation value of  the Svetlichny operator (which is bounded from above by the inequality) for the two inequivalent subclasses of pure three-qubit states: the GHZ-class and the W-class. We show that the maximum for the GHZ-class states reduces to Mermin's inequality [Phys. Rev. Lett. {\bf 65}, 1838 (1990)] modulo a constant factor, and although it is a function of the three tangle and the residual concurrence, large number of statesdon't violate the inequality. We further show that by design SI is more suitable as a measure of genuine tripartite nonlocality between the three qubits in the the W-class states, and the maximum is a certain function of the bipartite entanglement (the concurrence) of the three reduced states, and only when their certain sum attains a certain threshold value, they violate the inequality.
\end{abstract}



\maketitle
It is precisely the nonlocality (NL) of quantum correlations which gives an advantage to quantum mechanics over classical theories for certain information processing tasks~\cite{Nielsen2000}. The NL not only distinguishes quantum mechanics from a classical theory, but as far as the speedup of a quantum computational task is concerned, the problem of quantifying the NL of a multipartite quantum state is indispensable. The correlations between outcomes of measurements on two or more spatially separated subsystems are said to be {\it nonlocal}, if they cannot be simulated with shared randomness (which is commonly referred as {\it hidden variables}) without communication, {\it i.e.}, one cannot give a classical model which explains the correlations~\cite{Bell}. One defines an appropriate Bell-type inequality (BTI), since it gives an upper bound on correlations which are consistent with any local hidden-variable, or local-{\it realistic}, theory~\cite{Bell}. Thus, the amount of violation of such an inequality by an entangled state  is said to be a {\it measure} of the NL of the correlations between the subsystems.

It is known that all $2$-qubit pure states violate the bipartite BTI, known as the CHSH inequality~\cite{Clauser69}, and the violation increases as the entanglement of the state increases~\cite{Gisin91}. The extension of this result to $3$-qubit pure states is nontrivial. For instance, Mermin's tripartite BTI is based on {\it absolute} local realism~\cite{Mermin}, {\it i.e.}, it is derived on the assumption that all the three qubits are locally but realistically correlated, and its violation is supposed to be a measure of {\it irreducible} (genuine) tripartite nonlocal correlations between the qubits. However, the bi-separable states do violate the inequality~\cite{Collins02}. This motivated Svetlichny to formulate a {\it hybrid} nonlocal-local  realism based inequality~\cite{Svetlichny87}: a stronger kind of inequality  for a three-qubit  system  where two of  the  qubits  are assumed to be non-locally correlated, but they  are locally correlated to the third, and one takes an ensemble  average over all such possible combinations.  Thus, by  construction, the violation  of Svetlichny's inequality (SI) is a signature of {\it genuine} tripartite nonlocality, but it is not a necessary requirement. In this letter we use Svetlichny's inequality (SI) to quantify genuine tripartite nonlocality of the following two subclasses of $3$-qubit pure states in terms of their genuine tripartite entanglement~\cite{Ghose09}-- the GHZ-class states
\begin{equation}
|\psi_{gs}\rangle=\cos\theta|000\rangle+\sin\theta|11\rangle\Big\{\cos\theta_3|0\rangle+\sin\theta_3|1\rangle\Big\}\;,
\label{gslice}
\end{equation}
and the W-class states
\begin{equation}
\label{wstate}
|\psi_{w}\rangle=\alpha|001\rangle+\beta |010\rangle+\gamma|100\rangle\;,
\end{equation}
where $\alpha$, $\beta$, and $\gamma$ are real.

{\it The monogamy:} The reason which complicates the study of nonlocality of $3$-qubit pure states is that the entanglement in the two classes are {\it inequivalent}~\cite{Dur00}. The difference can be quantified by a measure of genuine tripartite entanglement called the {\it three-tangle}~\cite{Coffman00}:
\begin{equation}
\tau(\psi) = {\cal C}_{1(23)}^2-{\cal C}_{12}^2-{\cal C}_{13}^2\;,
\label{tau}
\end{equation}
which is invariant under all permutations of subsystem indices; and where the {\it concurrence} ${\cal C}_{1(23)}^2$ is bipartite entanglement between qubit  $1$  and  qubits $2$-$3$,  and ${\cal C}_{12}^2$ is the concurrence of the reduced state $\rho_{12}$~\cite{Coffman00}. $\tau\ge 0$ characterizes the generalized GHZ state, whereas $\tau=0$ for {\it all} the W-class states. Since $\tau$ is an entanglement {\it monotone}, hence the in-equivalence~\cite{Dur00}. The difference arises in the way the bipartite entanglement is {\it distributed} among the qubits, {\it i.e}, the concurrences are constrained by the monogamy inequality~\cite{Coffman00}:
\begin{equation}
{\cal C}_{1(23)}^2\geq{\cal C}_{12}^2+{\cal C}_{13}^2\;,
\label{monogamy}
\end{equation}
which is saturated by W-class states, while the difference ${\cal C}_{1(23)}^2 - {\cal C}_{12}^2-{\cal C}_{13}^2$ is maximized by the GHZ-class states. This implies that the W-class states is determined by the concurrences of the three reduced states (modulo local unitaries), and bigger the sum of the concurrences, more its tripartite entanglement; in contrast, the GHZ-class states are fixed by the tangle and the {\it residual} concurrences, where the latter reduces the genuine tripartite entanglement of the state~\cite{Linden02}.

{\it Svetlichny's inequality:} Let the measurements by observers be spin projections onto unit vectors:  $A={\vec{\sigma_1}}\cdot{\vec{a}}$ or $A^\prime={\vec{\sigma_1}}\cdot{\vec{a^\prime}}$ on qubit $1$, $B={\vec{\sigma_2}}\cdot{\vec{b}}$ or $B^\prime={\vec{\sigma_2}}\cdot{\vec{b^\prime}}$ on qubit $2$, and $C={\vec{\sigma_3}}\cdot{\vec{c}}$ or $C^\prime={\vec{\sigma_3}}\cdot{\vec{c^\prime}}$ on the third qubit. If a theory is consistent with the hybrid nonlocal-local realism, then the quantum prediction for any $3$-qubit state $|\Psi\rangle$ is bounded by Svetlichny's inequality~\cite{Svetlichny87}:
\begin{equation}
|\langle\Psi|S|\Psi\rangle|\equiv S(\Psi)\leq 4\;.
\label{SV}
\end{equation}
where the Svetlichny's operator $S$  is defined as
\begin{equation}
S=A(DC+D^\prime C^\prime)+A'(D^\prime C-DC^\prime) =M+M^{\prime}\;,
\label{soperator}
\end{equation}
where $D=B+B'$ and $D'=B-B'$, and $\langle M\rangle\leq 2$  and $\langle M^\prime\rangle\leq 2$ are Mermin's inequalities~\cite{Cereceda02}.

Note that $S$ can be further simplified by defining ${\vec{b}}+{\vec{b^\prime}}=2{\vec{d}}\cos t$ and ${\vec{b}}-{\vec{b^\prime}}=2{\vec{d^\prime}}\sin t$, which implies
\begin{equation}
{\vec{d}}\cdot{\vec{d^\prime}}=\cos\theta_d\cos\theta_{d^\prime}+\sin\theta_d\sin\theta_{d^\prime}\cos(\phi_d - \phi_{d'})=0.
\label{orthogonal}
\end{equation}
Now by setting $D=\vec{d}\cdot \vec{\sigma}_2$ and $D'=\vec{d'}\cdot \vec{\sigma}_2$, gives
\begin{eqnarray}
S(\Psi)&=& 2|\cos t\langle ADC \rangle+ \sin t\langle AD'C'\rangle\\
&-&\cos t\langle A'DC'\rangle +\sin t\langle A'D'C\rangle|\nonumber\\
& \leq &  2\Big | \Big\{\langle ADC\rangle^2
+\langle AD'C' \rangle^2\Big\}^{\frac{1}{2}}\nonumber\\
&+& \Big\{\langle A'DC' \rangle^2 + \langle A'D'C\rangle^2
\Big\}^{\frac{1}{2}} \Big |,
\label{inequality0}
\end{eqnarray}
where we have used the fact that
\begin{equation}
\label{maxtheta}
x\cos\theta+y\sin\theta\leq (x^2+y^2)^{\frac{1}{2}}\;,
\end{equation}
the equality results when $\tan\theta=y/x$. The following
\begin{equation}
\label{maxtheta2}
x\sin^2\theta+y\cos^2\theta \leq \left \{
\begin{array}{lr}
y, \; & x \leq y\;\\
x, \;& x \geq y\;,
\end{array}
\right.
\end{equation}
will be useful later; the first inequality is realized when $\theta=0$,
and $\theta=\pi/2$ gives the second. In the next two sections, we obtain the maximum value of the expectation value Svetlichny's operator, $S_{\text{max}}(\psi)$, with respect the GHZ-class states $|\psi_{gs}\rangle$~(\ref{gslice}) and the W-class states $|\psi_w\rangle$~(\ref{wstate}).

{\it The GHZ-class states:} Let $P=(1-2\sin^2\theta\sin^2\theta_3)$,
$Q=(\sin^2\theta\sin2\theta_3)$,
$\cos\phi_{adc}=\cos(\phi_a+\phi_d+\phi_c)$,                        and
$\cos\phi_{ad}=\cos(\phi_a+\phi_d)$, then the first term $\langle ADC \rangle$  in
(\ref{inequality0}) with respect to $|\psi_{gs}\rangle$ can be expressed as
\begin{eqnarray}
&&\cos\theta_a\cos\theta_d\big\{P\cos\theta_c+Q\cos\phi_c\sin\theta_c\big\} +\Big\{\sin2\theta\sin\theta_a \nonumber\\
&&\sin\theta_d\big\{\cos\theta_3\cos\phi_{ad}\cos\theta_c
+\sin\theta_3\cos\phi_{adc}\sin\theta_c\big\}\Big\}\;,
\label{gsinequality01}
\end{eqnarray}
which when maximized with respect to $(\phi_d - \phi_{d^{\prime}})$ by using (\ref{orthogonal}) and  considering $\theta_{d^{\prime}}$, $\phi_d$, and $(\phi_d - \phi_{d^{\prime}})$ to be independent variables, one obtains $(\phi_d - \phi_{d^{\prime}})= 0$ and $\theta_d=\frac{\pi}{2}$. The iterative maximization of the Mermin operator (\ref{soperator}) using inequalities (\ref{maxtheta}) and (\ref{maxtheta2}) is summarized below:
\begin{widetext}
\begin{eqnarray}
M&=&2\Big\{ \langle ADC \rangle^2 + \langle AD^{\prime}C^{\prime} \rangle^2 \Big\}^\frac{1}{2}\\
&\leq&2\Big\{\sin^2\theta_a\sin^22\theta\big\{(\cos\theta_3\cos\phi_{ad}\cos\theta_{c}+\sin\theta_3\cos\phi_{adc}\sin\theta_c)^2 + (\cos\theta_3\sin\phi_{ad}\cos\theta_{c^{\prime}}+\sin\theta_3\sin\phi_{ad{c^{\prime}}}\sin\theta_{c^{\prime}})^2\big\}\nonumber\\
&+& \cos^2\theta_a(P\cos\theta_{c^{\prime}}+Q\cos\phi_{c^{\prime}}\sin\theta_{c^{\prime}})^2\Big\}^{\frac{1}{2}}\label{mermin0}\\
&\leq & \left\{ \begin{array}{lr}
2\sin2\theta\big\{(\cos\theta_3\cos\phi_{ad}\cos\theta_{c}+\sin\theta_3\cos\phi_{adc}\sin\theta_c)^2 + (\cos\theta_3\sin\phi_{ad}\cos\theta_{c^{\prime}}+\sin\theta_3\sin\phi_{ad{c^{\prime}}}\sin\theta_{c^{\prime}})^2\big\}^{\frac{1}{2}}\\
2(P\cos\theta_{c^{\prime}}+Q\cos\phi_{c^{\prime}}\sin\theta_{c^{\prime}})
\end{array}\right.\label{mermin1}\\
&\leq & \left\{ \begin{array}{lr}
2\sin2\theta\big\{(\cos^2\theta_3\cos^2\phi_{ad}+\sin^2\theta_3\cos^2\phi_{adc}) + (\cos^2\theta_3\sin^2\phi_{ad}+\sin^2\theta_3\sin^2\phi_{ad{c^{\prime}}})\big\}^{\frac{1}{2}}\\
2(P^2+Q^2\cos^2\phi_{c^{\prime}})^{\frac{1}{2}}
\end{array}\right.\label{mermin2}\\
&\leq & \left\{ \begin{array}{lr}
2\sin2\theta\sqrt{1 + \sin^2\theta_3}\\
2\sqrt{P^2+Q^2} = 2(1-\sin^22\theta\sin^2\theta_3)^{\frac{1}{2}}\;,
\end{array}\right.\label{mermin3}
\end{eqnarray}
\end{widetext}
Maximization is over $\theta_{d^{\prime}}$ in (\ref{mermin0}), $\theta_a$ in (\ref{mermin1}), and $\theta_c$ and $\theta_{c^{\prime}}$ in (\ref{mermin2}). Equations (\ref{const1}) and (\ref{const2}) are a particular instance of the constraints that have to be satisfied for the top and bottom inequalities in (\ref{mermin3}) respectively,
\begin{eqnarray}
&&\left\{\begin{array}{lr} 
\theta_{d^{\prime}}=\frac{\pi}{2} \: ; \:\theta_a=\frac{\pi}{2}\: ; \:\theta_c=\theta_3\: ; \:\theta_{c^{\prime}}=\frac{\pi}{2}\\
\phi_{ad}=0 \:;\:\phi_{adc} =0\:;\:\phi_{adc^{\prime}} =\frac{\pi}{2}
\end{array}\right.\label{const1}\\
&&\theta_{d^{\prime}}=0 \: ; \:\theta_a=0 \: ;\theta_c=\frac{\pi}{2} \: ; \:\phi_{c^{\prime}}=0
\label{const2}
\end{eqnarray}
By symmetry in (\ref{inequality0}), $M^{\prime}$ is obtained by taking $A\leftrightarrow A^{\prime}$ and $C\leftrightarrow C^{\prime}$, and satisfying similar constraints to (\ref{const1}) and (\ref{const2}). More importantly, both sets of constraints can be {\it matched}; this implies that as far as the GHZ-class states is concerned, SI reduces to Mermins' inequality, modulo the constant value of $2$, which ensures that the violation of SI is sufficient to detect genuine tripartite nonlocality.

Equation~(\ref{mermin3}) implies that $|\psi_{gs}\rangle$ is
\begin{equation}
S_{\text{max}}(\psi_{gs})=\left\{
\begin{array}{lr}
4\sqrt{1-\tau}\;, & 3\tau+C^2_{12}\leq 1\\
4\sqrt{C^2_{12}+2\tau}\;, & 3\tau+C^2_{12}\geq 1\\
\end{array}
\right.\label{gsfinal3}
\end{equation}
where, as discussed earlier, the entanglement of $|\psi_{gs}\rangle$ is fixed by its tangle:
\begin{equation}
\tau(\psi_{gs})=\sin^2{2\theta}\sin^2\theta_3\;,
\end{equation}
and the residual concurrence of $\rm{tr}_3(|\psi_{gs}\rangle\psi_{gs}|)=\rho_{12}$:
\begin{equation}
C^2_{12}(\psi_{gs})=\sin^2{2\theta}\cos^2\theta_3\;,
\label{3concurrences}
\end{equation}
and, $C_{23}=C_{31}=0$.

\begin{figure}
\includegraphics[scale=0.32]{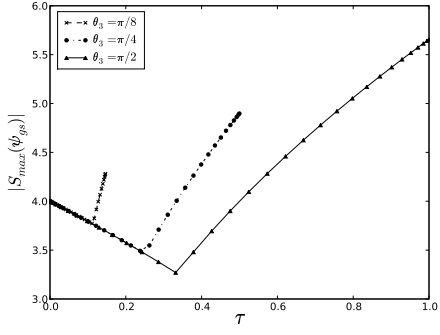}
\caption{Maximum of the Svetlichny operator for varying tangle $\tau$, for three values of $\theta_3 = \{\pi/8, \pi/4, \pi/2\}$.}
\label{fig:GHZ}
\end{figure} 

For instance, the first equality in (\ref{gsfinal3}) can be achieved by setting the measurement unit vectors as  ${\vec{a}}={\hat{x}}$,
${\vec{a^\prime}}={\hat{y}}$,
${\vec{b}}={\hat{x}}\cos{t}-{\hat{y}}\sin{t}$,
${\vec{b^\prime}}={\hat{x}}\cos{t}+{\hat{y}}\sin{t}$,
${\vec{c}}={\hat{z}}\cos\theta_3+{\hat{x}}\sin\theta_3$, and
${\vec{c^\prime}}={\hat{y}}$; and the set ${\vec{a}}={\hat{z}}$,
${\vec{a^\prime}}={\hat{z}}$,
${\vec{b}}={\hat{x}}\cos{t}+{\hat{z}}\sin{t}$,
${\vec{b^\prime}}={\hat{x}}\cos{t}-{\hat{z}}\sin{t}$,
${\vec{c}}=\hat{x}$,            and
${\vec{c^\prime}}={\hat{x}}$, where $\tan{t}=\sin\theta_3$,
attains the second in (\ref{gsfinal3}).
The behavior of $S_{\text{max}}(\psi_{gs})$ as a function of tripartite entanglement of $|\psi_{gs}\rangle$ is surprising (see Fig.\ref{fig:GHZ}). When the state is {\it tri-separable}, $\tau=C_{12}=0$, or {\it bi-separable}, $\tau=0$, $0<C_{12}\leq 1$, then as expected $S_{\text{max}}(\psi_{gs})=4$. In the region where the entanglement of  $|\psi_{gs}\rangle$ satisfy $3\tau+C^2_{12}\leq 1$, as the entanglement increases $S_{\text{max}}(\psi_{gs})$ monotonically decreases below the value of $4$ (this was also noted in Ref.~\cite{Ghose09}). The converse happens in the regime where  $3\tau+C^2_{12}\geq 1$, the value of $S_{\text{max}}(\psi_{gs})$ starts monotonically increasing as the entanglement increases; however only when ${C^2_{12}+2\tau}\geq 1$ do the states violate SI. Note in the latter region one expects the residual bipartite entanglement ${C^2_{12}}$ to decrease the maximum value, instead of increasing it. 

\begin{figure}
\includegraphics[scale=0.32]{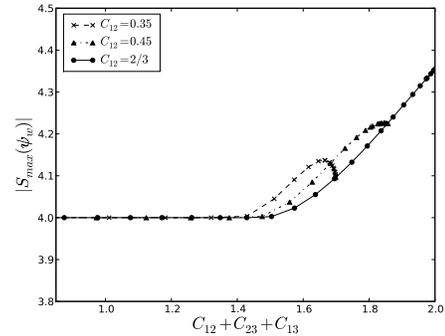}
\caption{Maximum of the Svetlichny operator for varying sum $(C_{12} + C_{23} + C_{31})\leq 2$, for three values of $C_{12} = \{0.35, 0.45, \frac{2}{3}\}$.}
\label{fig:Wclass}
\end{figure}

{\it The W-class states:}
For the W-class states it is convenient to obtain $S_{\text{max}}(\psi_{w})$ by simply adding all the eight terms involved in the Svetlichny's operator $S$, as all the terms in $S$ contribute differently. This, unlike the GHZ-class, makes SI significantly different from Mermin's inequality for the W-class states. Let $\cos\phi_{dc}=\cos(\phi_d-\phi_c)$, and likewise for similarly defined terms, then the term $\langle ABC\rangle$ in (\ref{soperator}) with respect to $|\psi_w\rangle$ can be expressed as
\begin{eqnarray}
&&\cos\theta_b(-\cos\theta_a\cos\theta_c+ C_{31}\sin\theta_a\sin\theta_c\cos\phi_{ac}) \label{winequality0}\\
&+&\sin\theta_b(C_{12}\cos\theta_a\sin\theta_c\cos\phi_{bc}+C_{23}\sin\theta_a\cos\theta_c\cos\phi_{ab} ),
\nonumber
\end{eqnarray}
where $C_{12}=2\alpha\beta,\;\; C_{23}=2\beta\gamma,\;\; C_{31}=2\gamma\alpha$ are the concurrences of the three reduced states of $|\psi_w\rangle$. Due to the inherent symmetry in (\ref{winequality0}), $S_{\text{max}}(\psi_{w})$ is achieved when all $\phi_i=0$. Now adding all the terms~(\ref{soperator}), one obtains for the expectation of Mermin operator:
\begin{widetext}
\begin{eqnarray}
\langle M\rangle &=& \frac{1}{4}\Big[(-1-C_{31}-C_{12}-C_{23})\Big\{\cos(\theta_a + \theta_b + \theta_{c^{\prime}}) + \cos(\theta_{a^{\prime}} + \theta_b + \theta_c) + \cos(\theta_a + \theta_{b^{\prime}} + \theta_c) - \cos(\theta_{a^{\prime}} + \theta_{b^{\prime}} + \theta_{c^{\prime}})\Big\}\nonumber\\
&+& (-1+C_{31}+C_{12}-C_{23})\Big\{\cos(\theta_a + \theta_b - \theta_{c^{\prime}}) + \cos(\theta_{a^{\prime}} + \theta_b - \theta_c) + \cos(\theta_a + \theta_{b^{\prime}} - \theta_c) - \cos(\theta_{a^{\prime}} + \theta_{b^{\prime}} - \theta_{c^{\prime}})\Big\}\nonumber\\
&+&(-1-C_{31}+C_{12}+C_{23})\Big\{\cos(\theta_a - \theta_b + \theta_{c^{\prime}}) + \cos(\theta_{a^{\prime}} - \theta_b + \theta_c) + \cos(\theta_a - \theta_{b^{\prime}} + \theta_c) - \cos(\theta_{a^{\prime}} - \theta_{b^{\prime}} + \theta_{c^{\prime}})\Big\}\nonumber\\
&+&(-1+C_{31}-C_{12}+C_{23})\Big\{\cos(\theta_a - \theta_b - \theta_{c^{\prime}}) + \cos(\theta_{a^{\prime}} - \theta_b - \theta_c) + \cos(\theta_a - \theta_{b^{\prime}} - \theta_c) - \cos(\theta_{a^{\prime}} - \theta_{b^{\prime}} - \theta_{c^{\prime}})\Big\}\Big]\nonumber
\end{eqnarray}
\end{widetext}
In the same fashion, one can find the expression for $\langle M^{\prime}\rangle$. The dependence on $\theta_i$'s can be suitably expressed by defining $\overline{\theta}_g =(\theta_g + \theta_{g^{\prime}})/2$, $\widetilde{\theta}_g= (\theta_{g^{\prime}} - \theta_g)/2$, $g\varepsilon\{a,b,c\}$. Allowing $\Sigma=(\widetilde{\theta}_a + \widetilde{\theta}_b + \widetilde{\theta}_c)$, and $\Sigma_g=\Sigma - 2\widetilde{\theta}_g$ one obtains,
\begin{widetext}
\begin{eqnarray}
S(\psi_{w}) &=& \frac{1}{2}\Big\{(-1-C_{31}-C_{12}-C_{23})\sin(\overline{\theta}_a + \overline{\theta}_b + \overline{\theta}_c)\{G - 2\sin(\widetilde{\theta}_a - \widetilde{\theta}_b -\widetilde{\theta}_c)\}\nonumber\\
&+&(-1+C_{31}+C_{12}-C_{23})\sin(\overline{\theta}_a + \overline{\theta}_b - \overline{\theta}_c)\{G - 2\sin(\widetilde{\theta}_a - \widetilde{\theta}_b +\widetilde{\theta}_c)\}\nonumber\\
&+&(-1-C_{31}+C_{12}+C_{23})\sin(\overline{\theta}_a - \overline{\theta}_b + \overline{\theta}_c)\{G - 2\sin(\widetilde{\theta}_a + \widetilde{\theta}_b -\widetilde{\theta}_c)\}\nonumber\\
&+&(-1+C_{31}-C_{12}+C_{23})\sin(\overline{\theta}_a - \overline{\theta}_b - \overline{\theta}_c)\{G - 2\sin(\widetilde{\theta}_a + \widetilde{\theta}_b +\widetilde{\theta}_c)\}\Big\}\nonumber\\
&=& \big\{-\sin\Sigma + \sin\Sigma_a + \sin\Sigma_b + \sin\Sigma_c \big\}+C_{13} \big\{\sin\Sigma + \sin\Sigma_a - \sin\Sigma_b + \sin\Sigma_c \big\}\nonumber\\
&+&C_{12} \big\{ \sin\Sigma - \sin\Sigma_a + \sin\Sigma_b + \sin\Sigma_c\big\}+ C_{23}\big\{\sin\Sigma + \sin\Sigma_a + \sin\Sigma_b - \sin\Sigma_c \big\}\label{second}\\
&\equiv& 4(p_1+p_2 C_{13}+p_3 C_{12} + p_4 C_{23})\;,
\label{Swclass}
\end{eqnarray}
\begin{equation}
G=\big\{\sin(\widetilde{\theta}_a + \widetilde{\theta}_b + \widetilde{\theta}_c) + \sin(\widetilde{\theta}_a + \widetilde{\theta}_b - \widetilde{\theta}_c) + \sin(\widetilde{\theta}_a - \widetilde{\theta}_b + \widetilde{\theta}_c) + \sin(\widetilde{\theta}_a - \widetilde{\theta}_b - \widetilde{\theta}_c) \big\}\;.
\end{equation}
\end{widetext}
and where the second equality (\ref{second}) is achieved when $\overline{\theta}_a=\overline{\theta}_b=\overline{\theta}_c={\pi/2}$. 
By symmetry, the global maximum of $S_{\text{max}}(\psi_{w})$ occurs when $C_{31}=C_{12}=C_{23}=2/3$, for which $\widetilde{\theta}_a = \widetilde{\theta}_b = \widetilde{\theta}_c = \widetilde{\theta}$. Then,
$|S_{\text{max}}(\psi_{w})| = \sin3\widetilde{\theta} + 5\sin\widetilde{\theta}$. The maximum occurs at $\widetilde{\theta}=54.736^{\circ}$ giving $S_{\text{max}}(\psi_{w}) = 4.354$~\cite{Cereceda02},
which can be obtained when the measurement directions are ${\vec{a}}={\vec{b}}={\vec{c}}=\hat{x}\cos\widetilde{\theta} + \hat{z}\sin\widetilde{\theta} $, and
${\vec{a^\prime}}={\vec{b^\prime}}={\vec{c^\prime}}=\hat{x}\cos\widetilde{\theta} - \hat{z}\sin\widetilde{\theta}$.
As expected, $S_{\text{max}}(\psi_{w})$ is $4$ for the tri-separable states (then only the first term survives in (\ref{Swclass})), and for the bi-separable states when the first two terms ($C_{13}\neq 0$) survives in (\ref{Swclass})). For arbitrary tripartite entangled states (see Fig. \ref{fig:Wclass}), the only states which violates SI  is when
$(p_1+p_2 C_{13}+p_3 C_{12} + p_4 C_{23})\geq 1$.

{\it Conclusion:} In this letter we quantified the genuine tripartite nonlocality of the subclass of $3$-qubit pure states, which can be generalized to all the pure states. Our main results showed that by construction SI is a suitable measure of tripartite nonlocality for the W-class states, for the GHZ-class states it reduces to Mermin's inequality, and gives counter intuitive results. A large number of states in both classes don't violate the inequality, which implies that perhaps the Svetlichny's kind of hybrid local-nonlocal theory is {\it too} strong by assumption, and thus can simulate the genuine tripartite correlations in such states. Elsewhere, we show how the inequality should be appropriately modified such that the resulting inequality completely quantifies the nonlocality of all the $3$-qubit pure states.

PR thanks S. Ghose for introducing him to SI. This letter is dedicated to the memory of Jharana Rani Samal.

\end{document}